\documentclass[12pt]{ussci}
\usepackage{graphicx}% Include figure files
\usepackage{graphicx}% Include figure files
\usepackage{dcolumn}% Align table columns on decimal point
\usepackage{bm}% bold math
\usepackage[dvipsnames]{xcolor}
\usepackage[utf8]{inputenc}
\usepackage[T1]{fontenc}
\usepackage{mathptmx}
\usepackage{nomencl}
\makenomenclature

\usepackage{blindtext}
\newcommand\papertopic{Turbulent Combustion}
\title{Generalized Joint Probability Density Function Formulation in Turbulent Combustion using DeepONet}

\author[1]{Rishikesh Ranade}
\author[2]{Kevin Gitushi}
\author[2, *]{Tarek Echekki}

\affil[1]{CTO Office, Ansys Inc, Canonsburg, PA, US}
\affil[2]{Mechanical Engineering, North Carolina State University, Raleigh, NC, US}
\affil[*]{Corresponding author: \email{techekk@ncsu.edu}}

\begin{document}
\maketitle

%====================================================================
\begin{abstract} % not to exceed 200 words
Joint probability density function (PDF)-based models in turbulent combustion provide direct closure for turbulence-chemistry interactions. The joint PDFs capture the turbulent flame dynamics at different spatial locations and hence it is crucial to represent them accurately. The joint PDFs are parameterized on the unconditional means of thermo-chemical state variables, which can be high dimensional. Thus, accurate construction of joint PDFs at various spatial locations may require an exorbitant amount of data. In a previous work, we introduced a framework that alleviated data requirements by constructing joint PDFs in a lower dimensional space using principal component analysis (PCA) in conjunction with Kernel Density Estimation (KDE). However, constructing the principal component (PC) joint PDFs is still computationally expensive as they are required to be calculated at each spatial location in the turbulent flame. In this work, we propose the concept of a generalized joint PDF model using the Deep Operator Network (DeepONet). The DeepONet is a machine learning model that is parameterized on the unconditional means of PCs at a given spatial location and discrete PC coordinates and predicts the joint probability density value for the corresponding PC coordinate. We demonstrate the accuracy and generalizability of the DeepONet on the Sandia flames, D, E and F. The DeepONet is trained based on the PC joint PDFs observed in flame E and yields excellent predictions of joint PDFs shapes at different spatial locations of flames D and F, which are not seen during training.
\end{abstract}

% Provide 2-4 keywords describing your research. Only abbreviations firmly
% established in the field may be used. These keywords will be used for
% sessioning/indexing purposes. Use \sep between each keyword.
\begin{keyword}
    Turbulent combustion\sep Joint PDF \sep Machine Learning \sep DeepONet
\end{keyword}

%====================================================================
\section{Introduction}

Modeling closure for turbulence-chemistry interactions presents an important challenge in turbulent combustion modeling \cite{Echekki2011}. Joint probability density function (PDFs) based approaches are commonly used to determine a direct closure for the turbulence-chemistry interactions in the governing equations and accelerate high fidelity simulations. Most turbulence-chemistry closure approaches fall under two categories, Flamelet-based and PDF-based~\cite{Pope2013}. The Flamelet approaches rely on the assumption of the PDF shape (e.g. $\beta$-PDF) and hence, the resulting closure models are limited to certain combustion modes and regimes. On the other hand, the PDF approaches calculate the joint PDF distributions. Although these methods are computationally costly, but they provide an accurate representation of the closure terms. 

In turbulent combustion simulations, the thermo-chemical space is high dimensional and this presents difficulties in the construction of joint PDFs. Hence, closure approaches for turbulent flames rely on effectively representing the thermo-chemical state in a lower dimensional set of variables. In many approaches, the lower-dimensional basis is established using physical variables, such as mixture fraction, progress variable, scalar dissipation etc., that efficient describe the turbulent flames. With the developments in machine learning and statistical methods, recent approaches have relied on available simulation or experimental data to extract a lower-dimensional basis from the thermo-chemical state using methods such as, Principal Component Analysis (PCA) \cite{Jolliffe}. PCA determines a low-dimensional basis for the data, such that the leading PCs represent the maximum explained variance in the data. In the context of turbulent combustion, PCA provides a linear projection to represent the entire thermo-chemical state by a significantly smaller set of parameters, known as principal components (PCs). PCA has been used in combustion for chemistry reduction (see for example \textcite{Vajda1985} ) as well as for the parameterization of the composition space \cite{Danby2006,Sutherland2009,Mirgolbabaei2013,Mirgolbabaei2014a,Mirgolbabaei2014b,Mirgolbabaei2015a,Mirgolbabaei2015b, Owoyele2017, Mirgolbabaei2015b, Ranade2019a, Ranade2019b}. 

The effective parameterization of the thermo-chemical state is an important first step in the computation of joint PDFs. As stated above, the Flamelet-based approaches presume a joint PDF shape in this lower dimensional manifold but PDF-based approaches rely on computation of the joint PDF. In the PDF-based approaches the joint PDF is computed mainly by PDF transport or by using data generated from lower-dimensional simulations, such as LEMLES \cite{Menon1996} and the LESODT \cite{Cao2008}. In these approaches, the turbulent flame is modeled stochastically on 1-D domains using the linear eddy model (LEM) \cite{Kerstein1989} and the one-dimensional turbulence (ODT) model \cite{Kerstein1999}. The scalar statistics collected at each spatio-temporal location are then used to calculate joint PDFs, which may be used to accelerate the full fidelity simulations. Recently, \textcite{Ranade2019a, Ranade2019b} introduced a novel turbulence-chemistry closure framework that used instantaneous experimental measurements to construct joint PDFs in a lower dimensional principal component (PC) manifold. The joint PC PDFs were subsequently used to determine the Favre means of thermo-chemical scalars and the PC source terms. In this work, the thermo-chemical parameterization was carried out using principal component analysis (PCA) \cite{Jolliffe}, while the joint PDFs were computed using kernel density estimation (KDE)~\cite{Bowman1997}.

Kernel density estimation (KDE)~\cite{Bowman1997} is a statistical method that can be used to construct arbitrary shapes for a joint PDF given discrete data at a given spatial position. In a KDE, a PDF can be expressed as a weighted sum of discrete kernel functions, $K$ (e.g. Gaussians) centered at discrete values of the parameters, say the PCs:
\begin{equation}\label{kde}
p\left( {{\bm\phi}} ; {\tilde{\bm\phi}} \right) = \frac{1}{n\ h} 
\sum_{i=1}^n K 
\left( \frac{{\bm{\phi}} - {\hat{\bm{\phi}}}_i}{h} \right).
\end{equation}
In this expression, $\bm{\phi}$ represents the vector of retained PCs, $K$ is the kernel function, $h$ is the so-called bandwidth, which controls the smoothing of the approximation, and $\hat{\phi}_i$ is the $i$th sample of $\phi$ values out of a total of $n$ samples. Such a distribution can be evaluated at a specific spatial point, which can be parameterized in terms of an unconditional mean for the PCs, $\tilde\phi$. Therefore, different shapes can be evaluated at different spatial positions. 

However, the KDE construction is very local and needs to be implemented at each spatial location in the space. Moreover, the computation of joint PDFs using KDE can be costly and require a large amount of data, as the lower-dimensional space increases in size. Hence, it is imperative to determine a ``generalized'' PDF that can be adopted for a wide range of unconditional means for the PCs or spatial position. In their recent work, de Frahan et al.~\cite{deFrahan2019} investigated the reconstruction of PDFs parameterized with the mixture fraction and the progress variable using three machine learning techniques, random forests, deep learning neural networks and conditional variational autoencoders. The machine learning techniques showed that the predictions o the joint PDFs yield improvements on $\beta$-function based presumed PDFs using DNS data.

The most recent work by \textcite{Lu2020} offers a potentially more powerful strategy to reconstruct joint PDFs for combustion data. In their approach, \textcite{Lu2020} have introduced the DeepONet network, which is designed to learn non-linear operators using sparse training data samples. The DeepONet networks consist of 2 separate input channels that provide different functional forms of the learned operators and their instances. The objective of this work is to present a generalizable model for the joint PDFs using the DeepONet network \cite{Lu2020} that can reconstruct joint PDFs at any given spatial location in a turbulent flame. The main function of the DeepONet is to replace the KDE-based PDF estimation, as used by \textcite{Ranade2019a, Ranade2019b}, and moreover, minimize the dependence on data requirements to estimate joint PDFs. The generalizable joint PDF model is trained on Sandia flames E and tested on all the Sandia flames D, E and F at different spatial locations in the flame. One of the main objectives of this work is to test the generalizability of the approach to unseen flames and flame conditions. The methodology of learning the PDF using DeepONet is described in Sec.~\ref{model}. Next, results are presented and discussed in Sec.~\ref{results}.

\section{Methodology} \label{model}
\subsection{PCA parameterization of composition space}

As stated earlier, PCA is a technique to linearly transform the vector of thermo-chemical scalars (e.g. temperature and species mass fractions) to uncorrelated scalars, the PCs:
\begin{equation}\label{pca}
    {\bf \phi} = {\bf A}^{\bf T} \ {\bf \psi},
\end{equation}
where $\bf\psi$ is the vector of $N$ thermo-chemical scalars normalized to yield comparable magnitudes using maximum and minimum values of these scalars: ${\bf\psi} = \left( T, Y_1, Y_2, \ldots, Y_{N-1}\right)$. The PCs vector $\bf\phi$ includes $N_{pc}$ principal components, ${\bf\phi} = \left( \phi_1, \phi_2, \ldots, \phi_{N_{pc}} \right)$, which account for a threshold percentage of the data variance. For example, the data from the Sandia flames~\cite{Barlow1998} and the Sydney flames~\cite{Meares2014,Meares2015,Barlow2015} require 2 to 3 PCs to capture the entire range of flow conditions and reaction scenarios (mixing, extinction and reignition) in these flames. The matrix $\bf A$ includes the leading $N_{pc}$ eigenvectors of the normalized thermo-chemical scalars covariance matrix.

The principal components (PCs) serve as a lower dimensional basis which can be used to compute the joint PDFs and conditional means, that may be used to reconstruct the unconditional Favre means of thermo-chemical scalars as well as source terms. The unconditional means for the $k$th thermo-chemical scalar or its source term can be expressed using the joint thermo-chemical scalars PDFs as follows:
\begin{equation}\label{uncondsource}
   {\overline {{\omega}_k \left( {\bf\phi} \right)}} = \int_{\bf\phi} \left<\omega_k \left| {\bf\phi} \right. \right> p\left( {\bf\phi} \right) d{\bf\phi}.
\end{equation}
\begin{equation}\label{uncondscalar}
   {\overline {{\psi}_k \left( {\bf\phi} \right)}} = \int_{\bf\phi} \left<\psi \left| {\bf\phi} \right. \right> p\left( {\bf\phi} \right) d{\bf\phi}.
\end{equation}
In this expression, $\left<\omega_k \left| {\bf\phi} \right. \right>$ is the mean of the thermo-chemical scalars source term and  $\left<\psi \left| {\bf\phi} \right. \right>$ is the mean of the thermo-chemical scalars conditioned on the PCs. $p\left( {\bf\phi}\right)$ is the joint PC PDF. Conditional means are constructed from data; they can be viewed as generalizations of flamelet libraries or FGMs where the parameterization is based on the PCs instead of prescribed parameters (e.g. mixture fraction, progress variable).

It is clear from the formulations in Eqs. \ref{uncondsource} and \ref{uncondscalar} that the accurate estimation of PDF at different locations in the flame is essential in predicting the means of source terms and thermo-chemical scalars. Next, we will discuss the DeepONet based joint PDF estimation.

\subsection{DeepONet based joint PDF}

\begin{figure*}[h!]
\begin{center}
\includegraphics[width=0.9\textwidth]{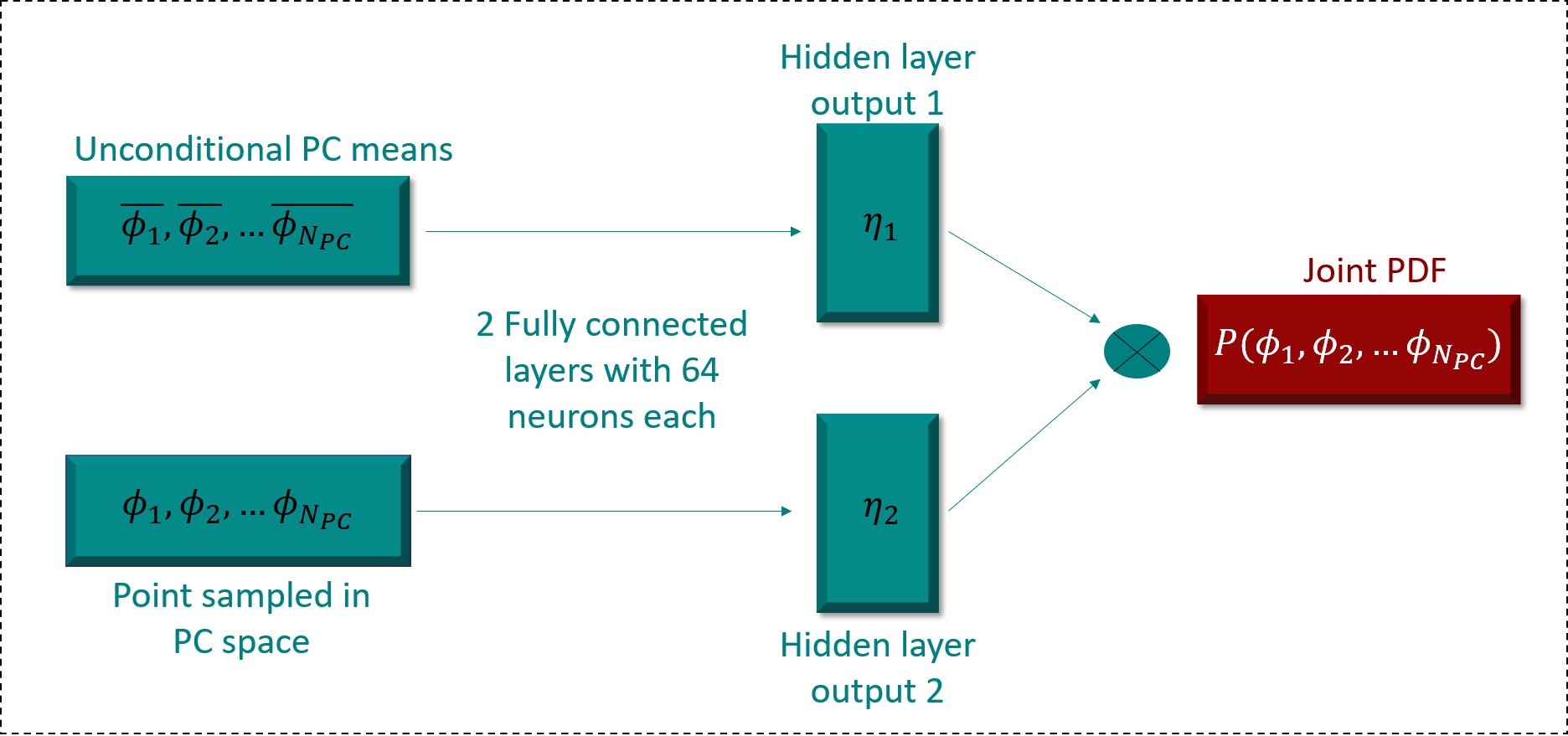}
\caption{\label{DeepONet} Illustration of a DeepONet for determining a PC's generalized PDF.}
\end{center}
\end{figure*}

The goal of the DeepONet is to predict a joint probability density function at any spatial location in a turbulent flame as a function of the PC state (derived from the thermo-chemical state) at that location. In the same spirit of general PDFs, the DeepONet ensures that a given PC state is conditioned with the unconditional mean and variances of PCs at a given location in the turbulent flame. However, as pointed out in our previous studies~\cite{Ranade2019a,Ranade2019b}, the use of the unconditional variances of PCs is not needed. This observation is primarily associated with the uncorrelated nature of PCs and the linear relations between thermo-chemical scalars and PCs.

Although ML-based methods to reconstruct parameterized PDFs have been proposed in the past~\cite{deFrahan2019}, the premise of the DeepONet networks is that they are designed to learn operators (in this case PDF shapes) that can relate prescribed inputs to desired outputs. The DeepONet architecture that corresponds to the present implementation is illustrated in Fig.~\ref{DeepONet}. It may be observed that the DeepONet allows for two channels at the input layer in order to accommodate different classes of inputs. The first input channel corresponds to the unconditional means of PCs at different spatial locations in the turbulent flame. Conversely, the second input channel is simply the PC state accessed by the turbulent flame. Both the input channels are passed through 2 fully connected layers with 64 neurons each. Finally, the outputs of these layers are multiplied element-wise and the vector sum represents the network output, which in this case is the probability density at a given PC state conditioned by an unconditional PC mean.

\subsection{Data generation for training}

The data generation process for training the DeepONet is outlined below.

\begin{enumerate}
    \item Given instantaneous thermo-chemical data at different spatial locations in the flame, a PCA is performed to determine the lower-dimensional PC representation of the instantaneous data at all the locations.
    \item Unconditional PC means are determined at each spatial location in the flame.
    \item Gaussian KDE is used to generate the joint PDFs at each spatial locations.
    \item The input data for DeepONet corresponds to the pair of instantaneous PC value and the unconditional PC mean at all locations in the flame. Similarly, the output data corresponds to the probability density value, determined from pre-computed KDE associated with the instantaneous PC conditioned by the unconditional PC mean. For example, if the lower dimensional basis is represented by 2 PCs, then each input sample for the DeepONet will correspond to $(\phi_1, \phi_2, \overline{\phi_1}, \overline{\phi_2})$ and the output sample will correspond to $P(\phi_1, \phi_2)$.
\end{enumerate}

\subsection{Training mechanics}

The DeepONet network is trained in TensorFlow~\cite{tensorflow}. Since DeepONets can learn operators from relatively sparse data sets, a sampling strategy is implemented to maintain the balance of low and high probabilities in the training set. A mean squared error loss is used to update the network weights and the training set is split such that 10\% is reserved for validation. The training data covers a sampled range of spatial positions in the turbulent flame; and hence the DeepONet network represents a generalized joint PDF, which can be recovered at any given point in the flame. 

\section{Results and Discussion} \label{results}
\subsection{Experimental conditions}

The dataset used in the following analysis is based on multiscalar measurements in the Sandia piloted jet flames by \textcite{Barlow1998}. The data set consists of 3 flames, D, E and F, characterized by different jet Reynolds numbers. Due to a difference in Reynolds numbers, these flames exhibit a varying degree of non-equilibrium effects such as flame extinction and reignition. Flame D has the smallest Reynolds numbers, followed by flame E and flame F. Other details related to the burner geometry and boundary conditions may be found in \textcite{Barlow1998}. The experimental data set corresponding to these flames consists of instantaneous measurements at $70-80$ axial and radial locations in these flames. The measured species corresponds to $T$, H$_2$, O$_2$, OH, CH$_4$, CO, CO$_2$ and H$_2$O. From a PCA analysis $3$ PCs represent $99\%$ of the explained variance and this is corroborated in a previous work carried out by the authors, \cite{Ranade2019a, Ranade2019b}. However, in this work we only retain the first $2$ PCs to learn the joint PDFs so as to build a proof of concept for the DeepONet approach. In future, we may extend this approach to using $3$ PCs.

Since, these flames exhibit a varying degree of non-equilibrium effects the joint PDF shapes observed at different locations of flame D can be significantly different from flames E and F. As a result in this work, we use data generated at a handful of spatial locations from flame E for training the DeepONet and test the generalizability of our approach by predicting joint PDFs on flames D and flame F.  
\begin{figure*}[h!]
\begin{center}
\includegraphics[width=0.5\textwidth]{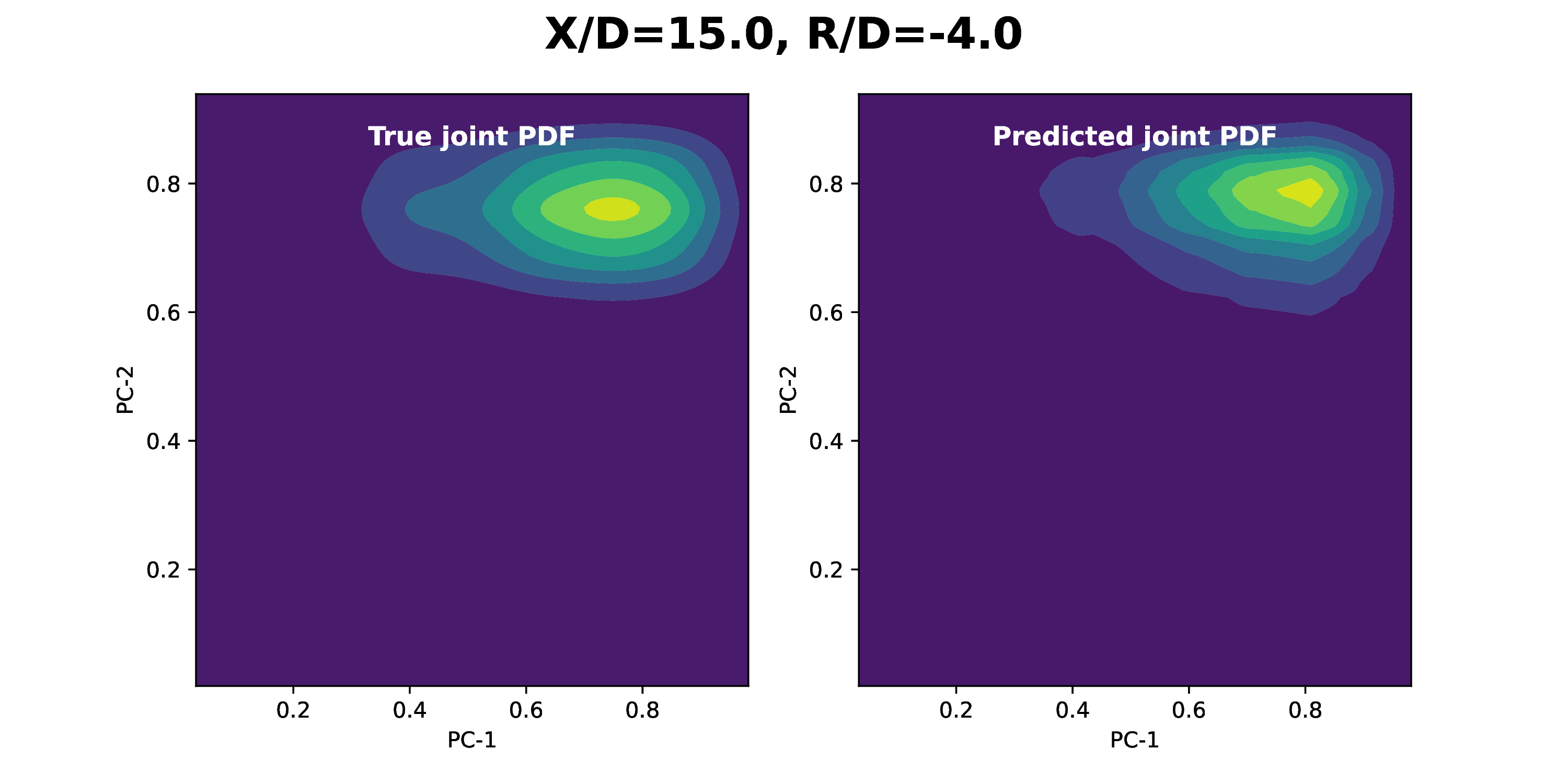}
\includegraphics[width=0.5\textwidth]{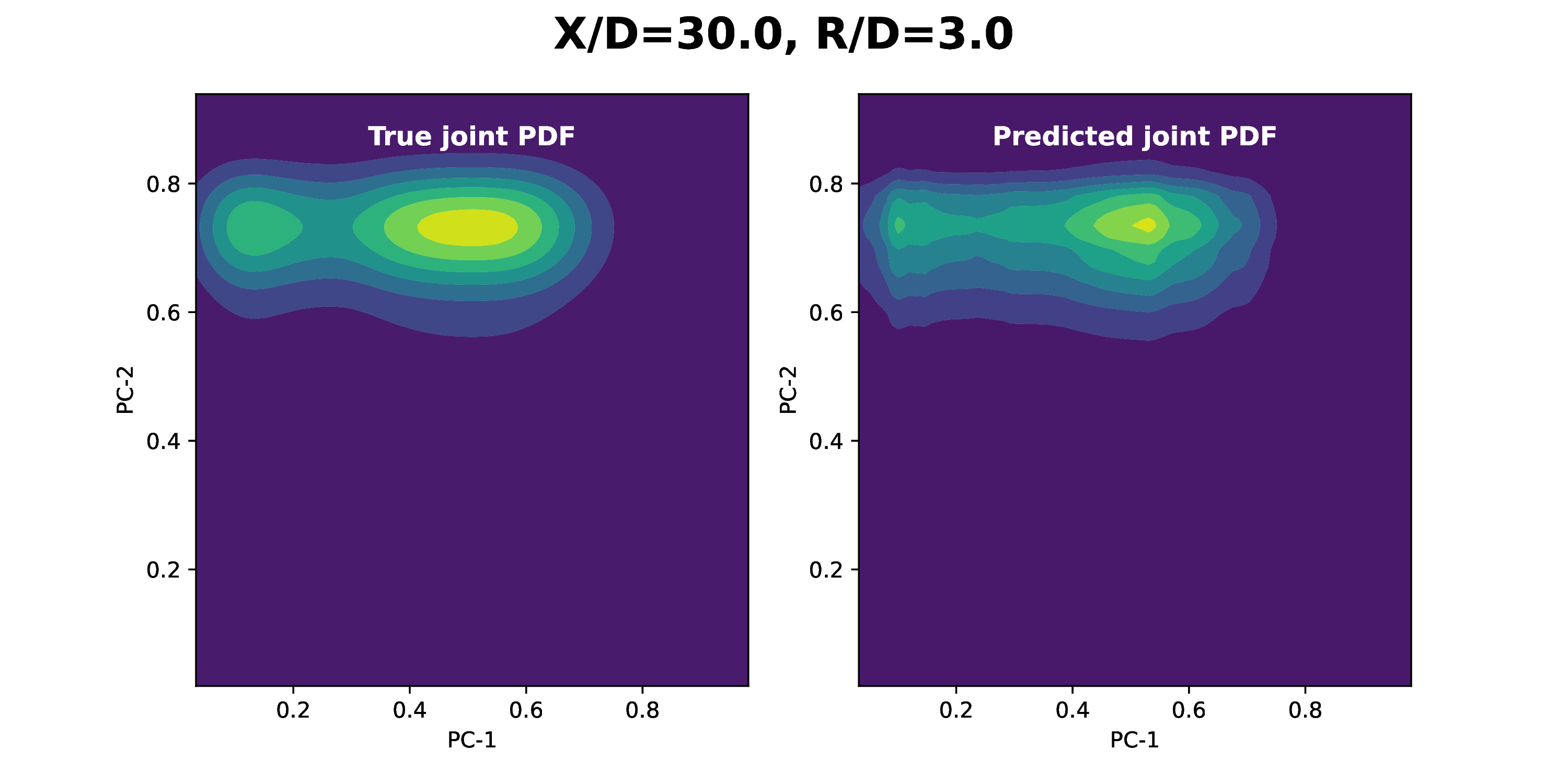}
\includegraphics[width=0.5\textwidth]{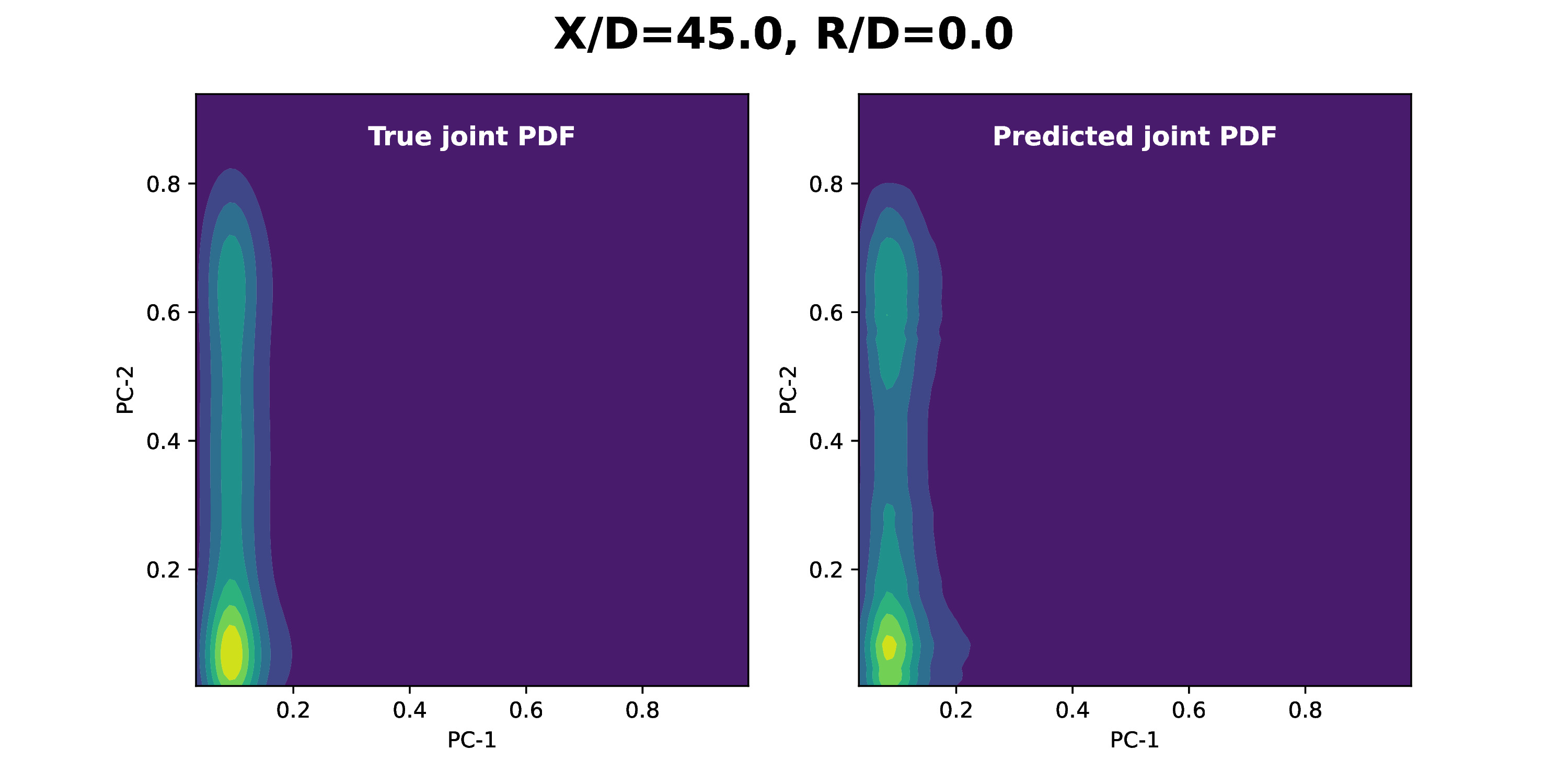}
\includegraphics[width=0.5\textwidth]{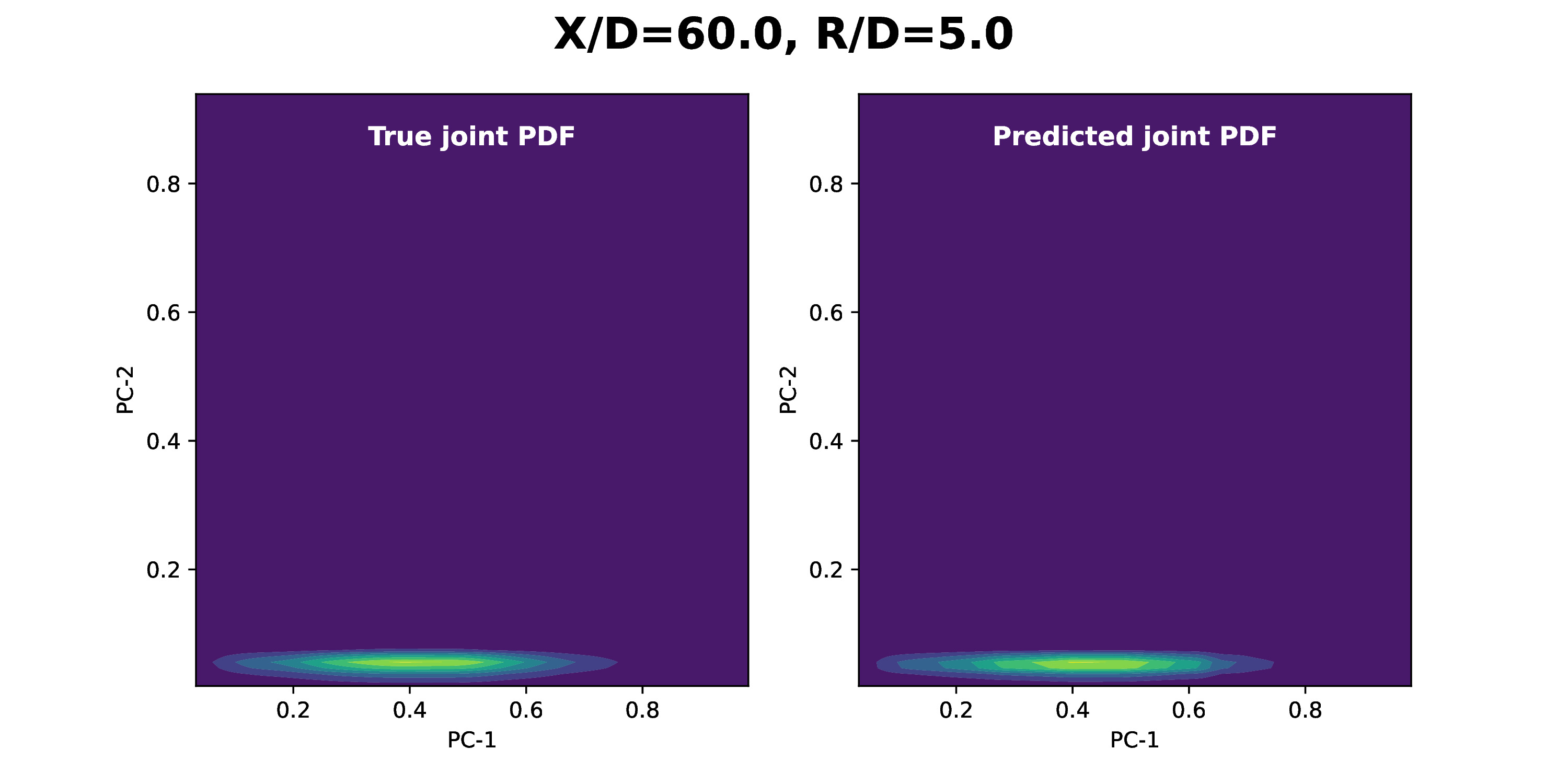}
\caption{\label{flamed} Comparison of contour plots between DeepONet vs KDE based joint PDFs for Sandia flame D}
\end{center}
\end{figure*}
\subsection{Contour comparisons}

\begin{figure*}[h!]
\begin{center}
\includegraphics[width=0.5\textwidth]{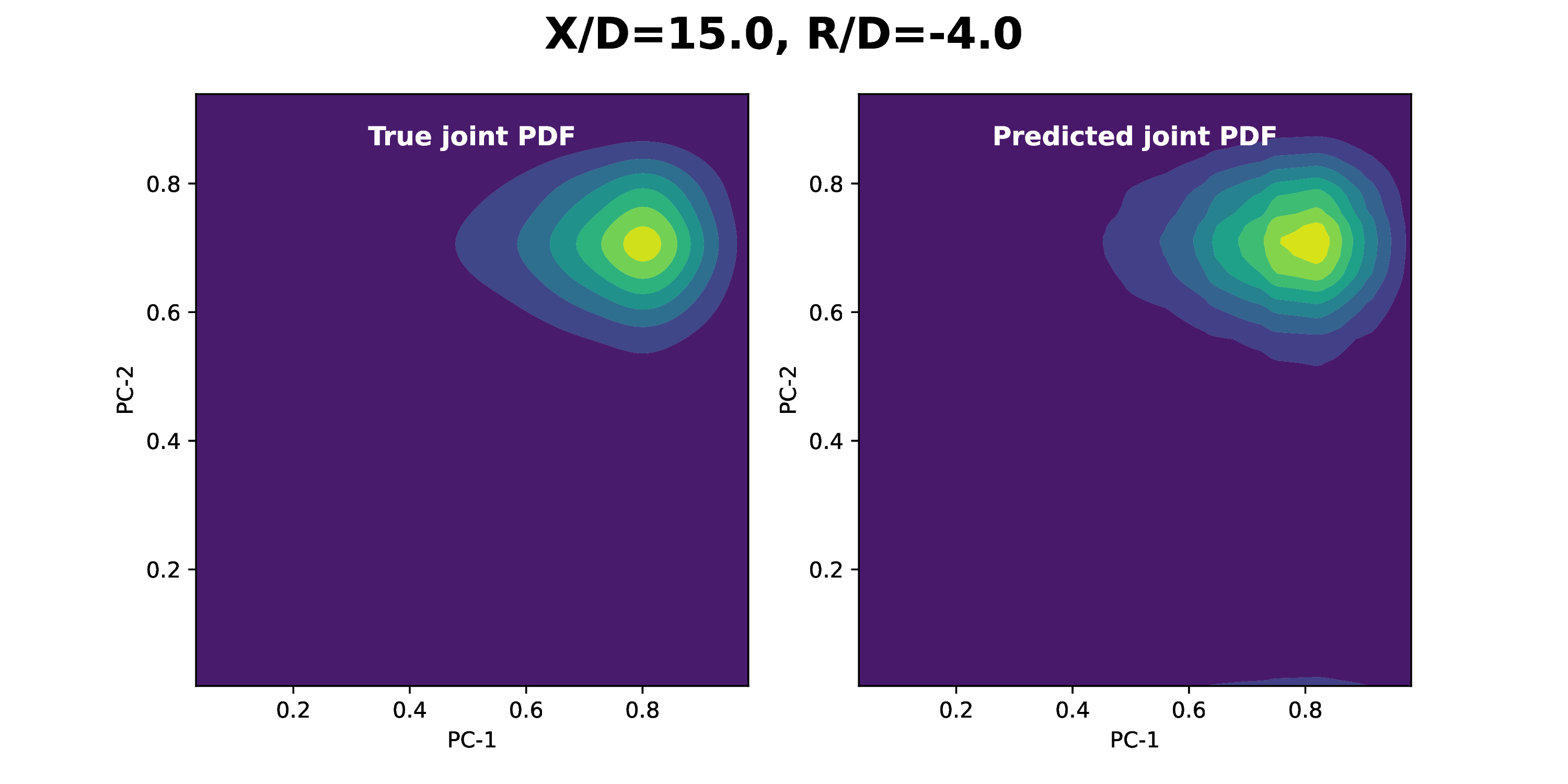}
\includegraphics[width=0.5\textwidth]{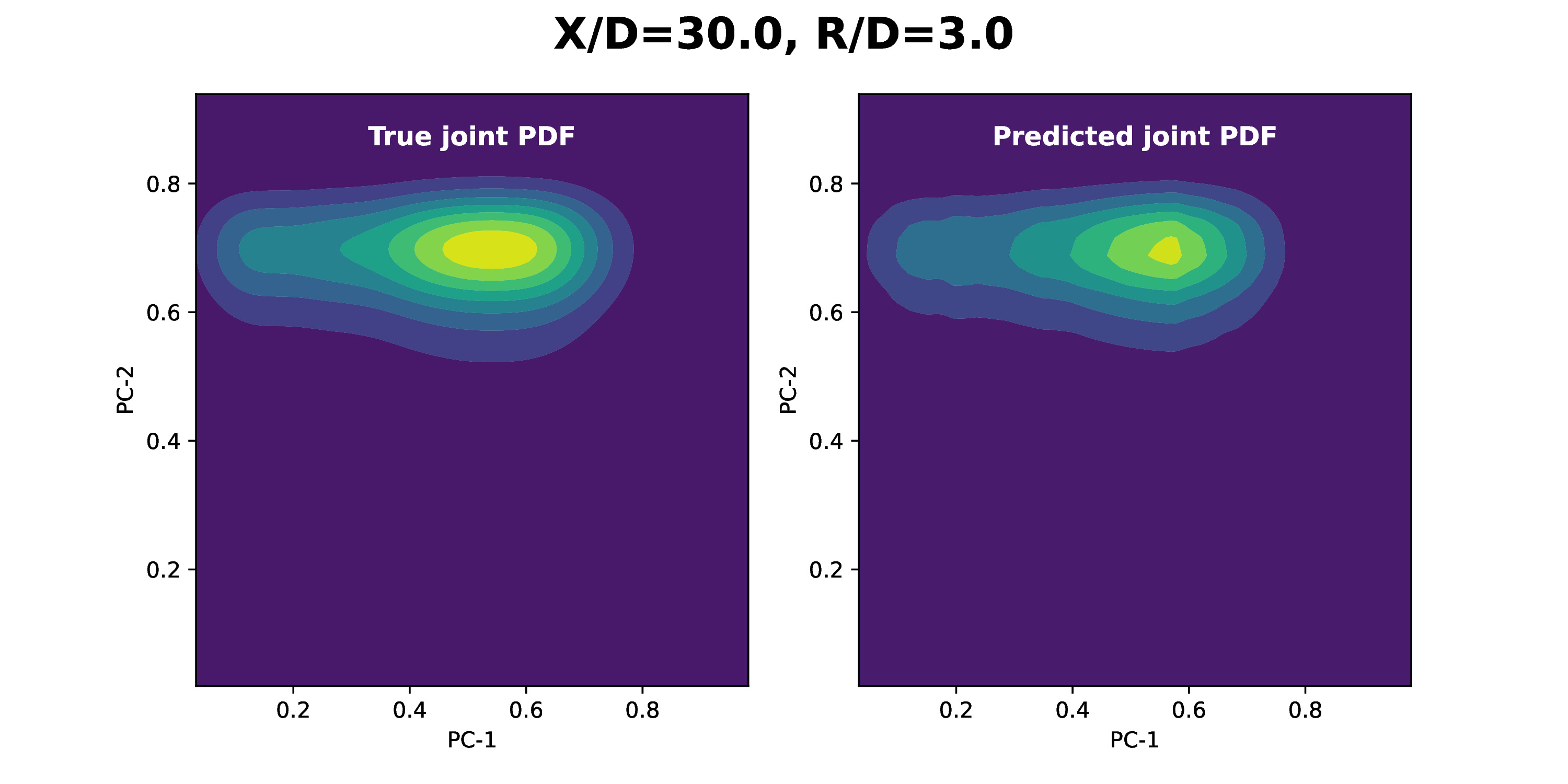}
\includegraphics[width=0.5\textwidth]{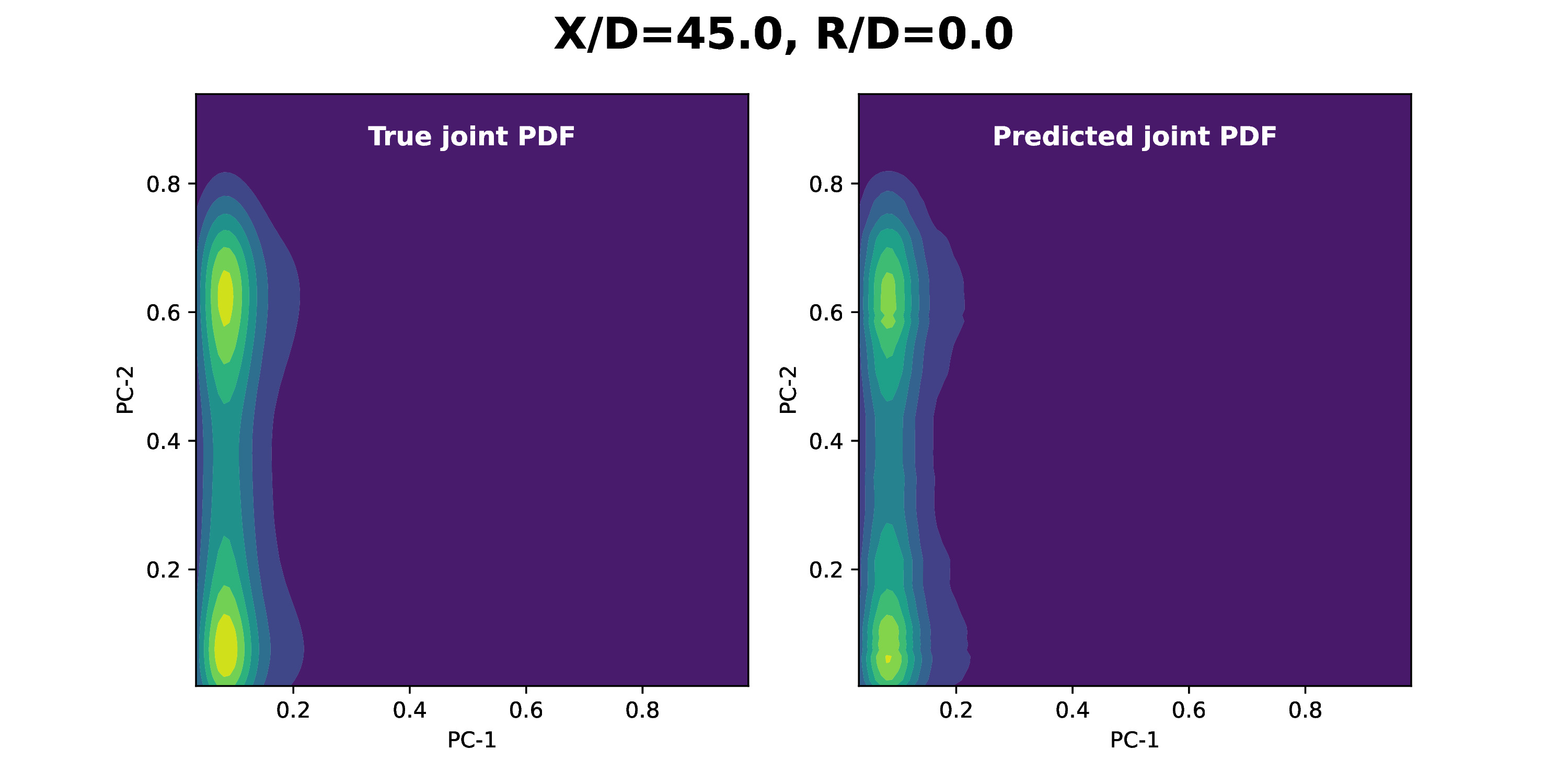}
\includegraphics[width=0.5\textwidth]{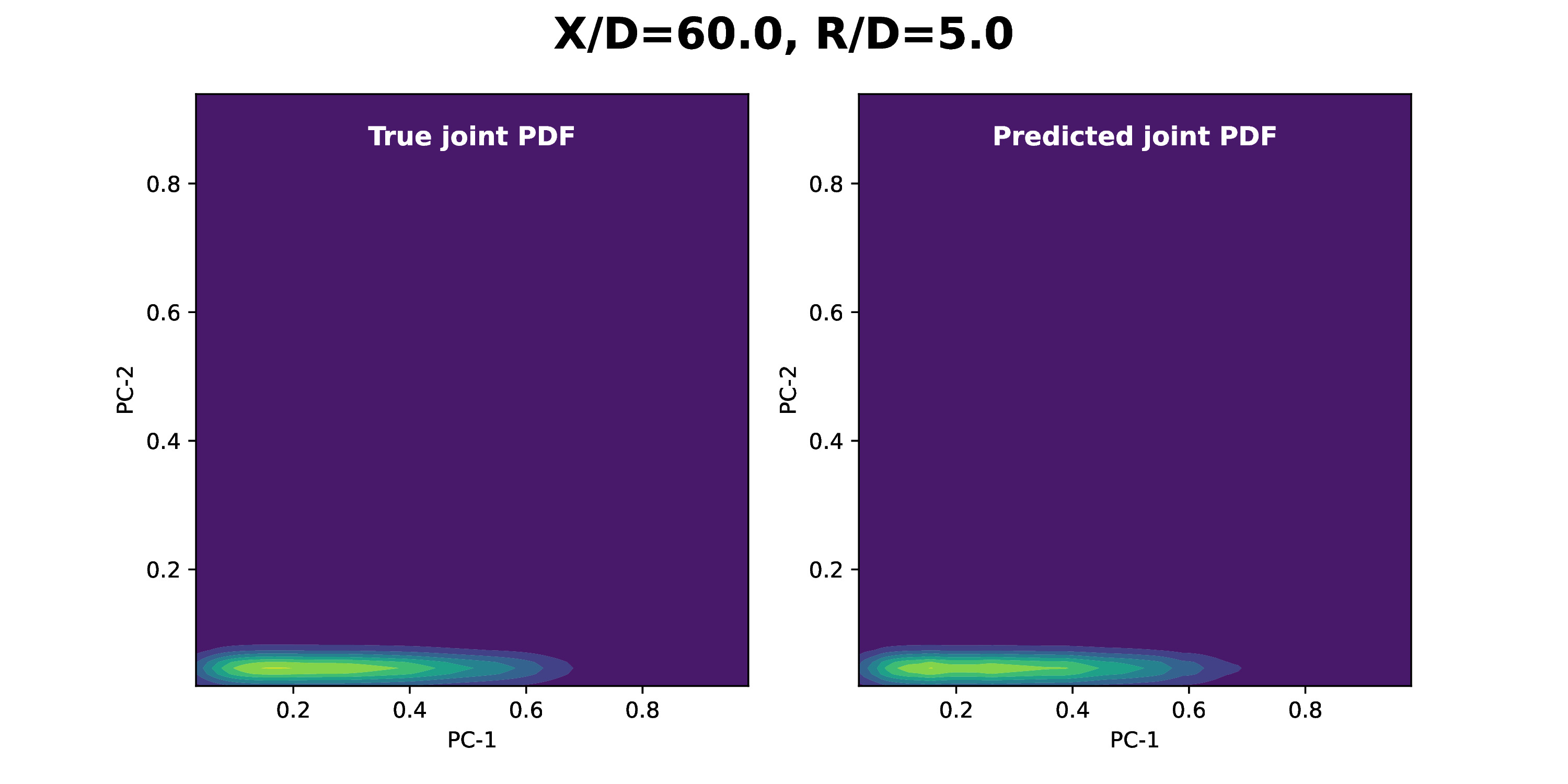}
\caption{\label{flamee} Comparison of contour plots between DeepONet vs KDE based joint PDFs for Sandia flame E}
\end{center}
\end{figure*}

\begin{figure*}[h!]
\begin{center}
\includegraphics[width=0.5\textwidth]{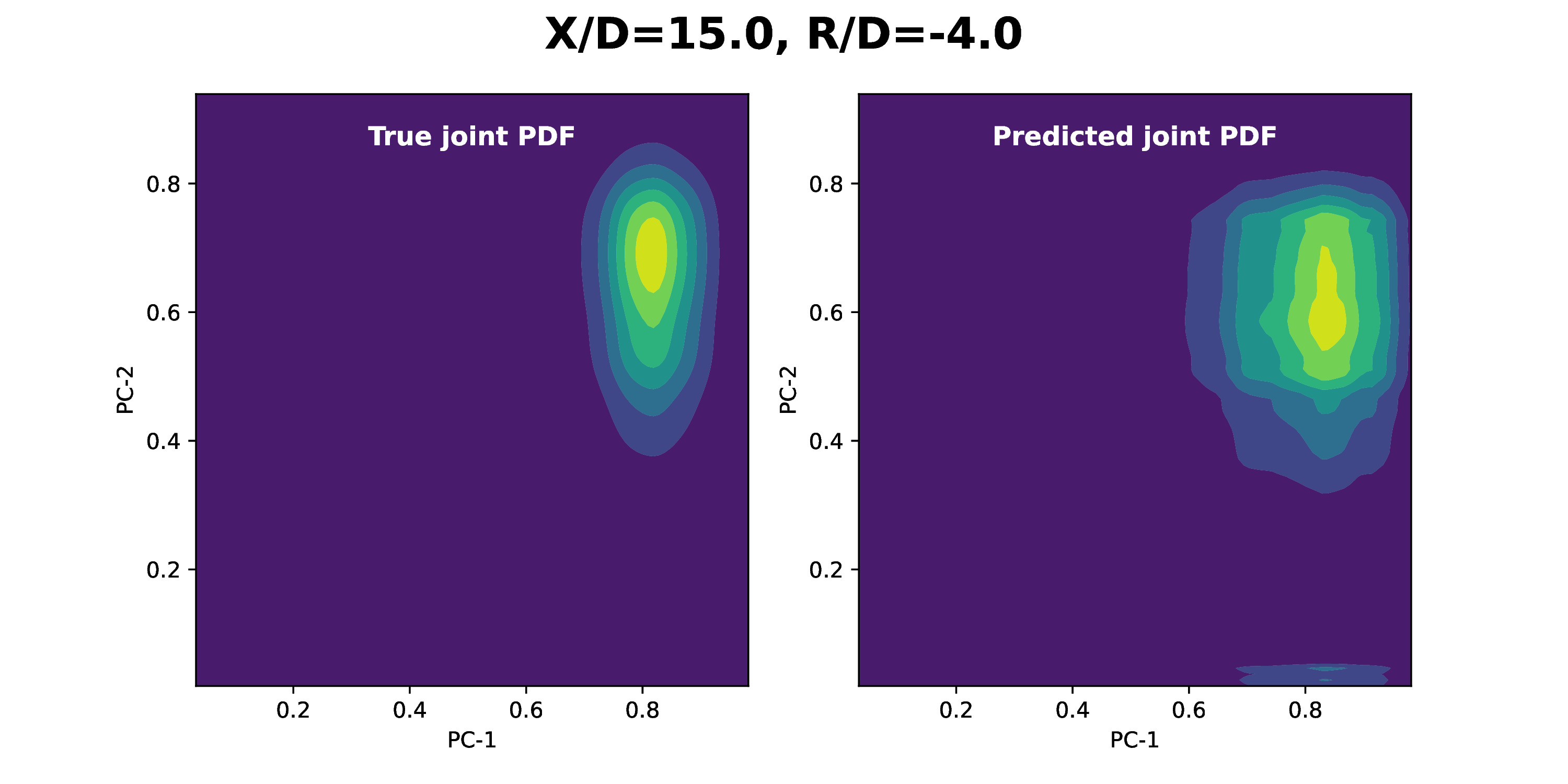}
\includegraphics[width=0.5\textwidth]{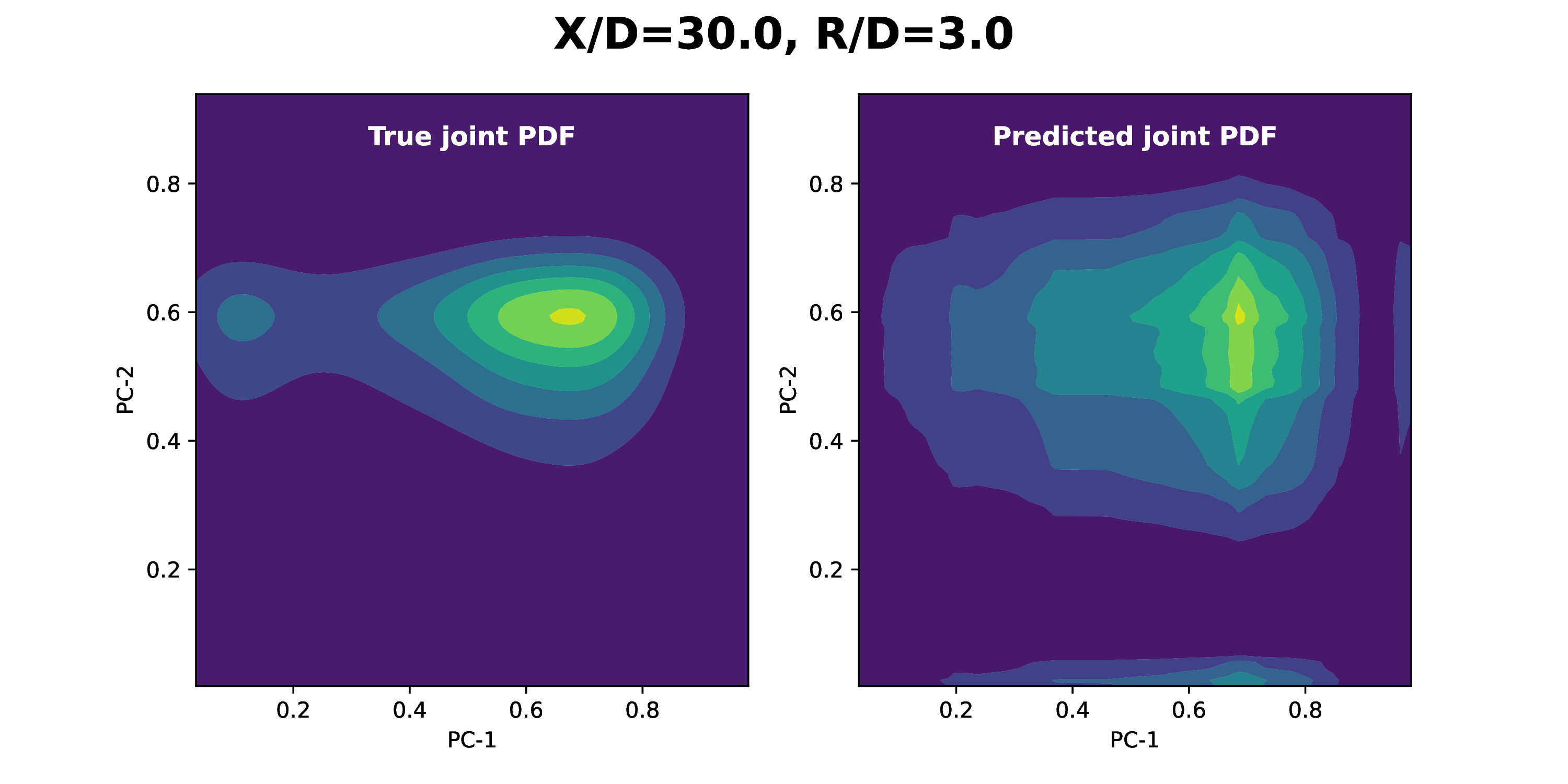}
\includegraphics[width=0.5\textwidth]{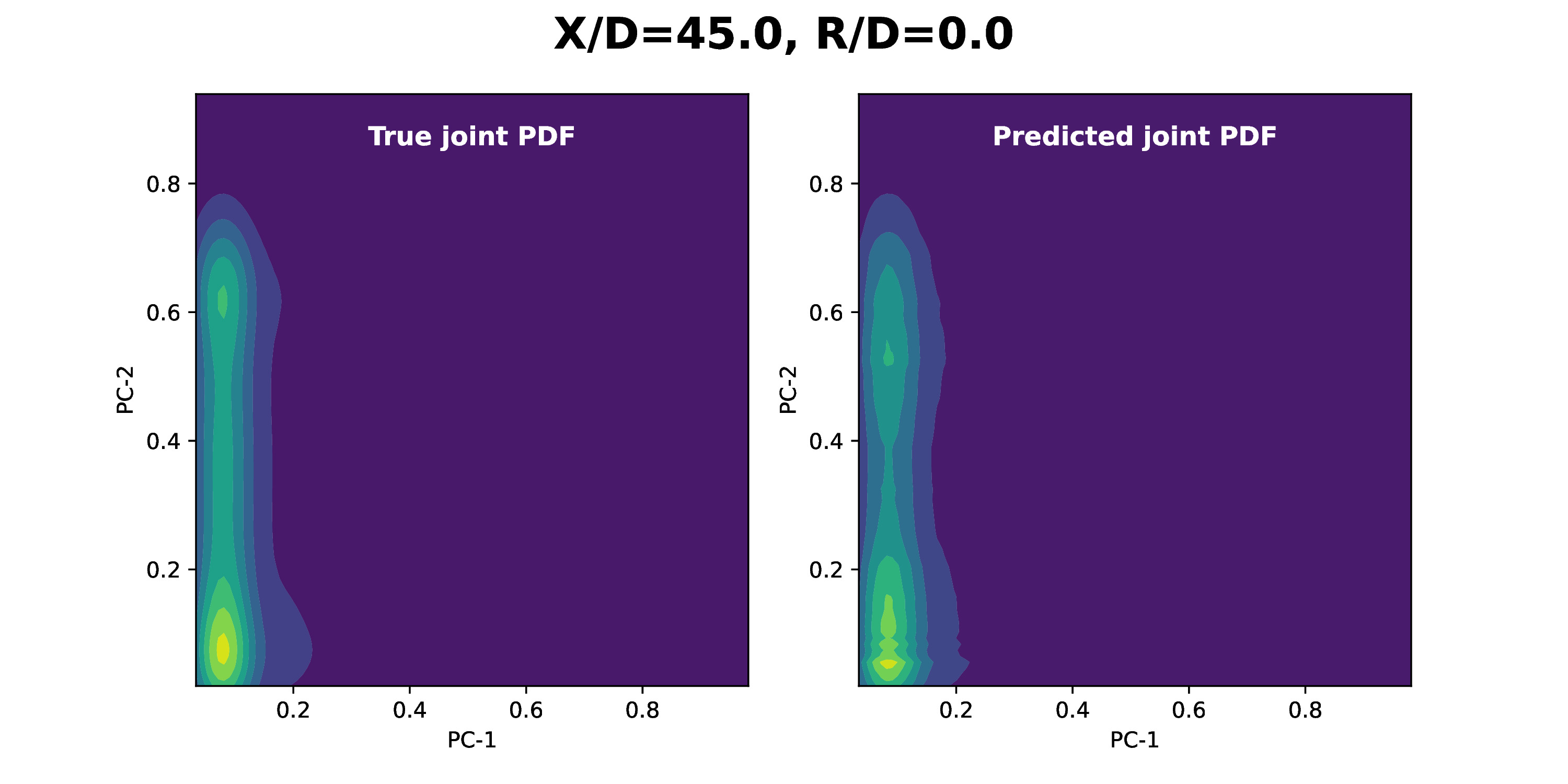}
\includegraphics[width=0.5\textwidth]{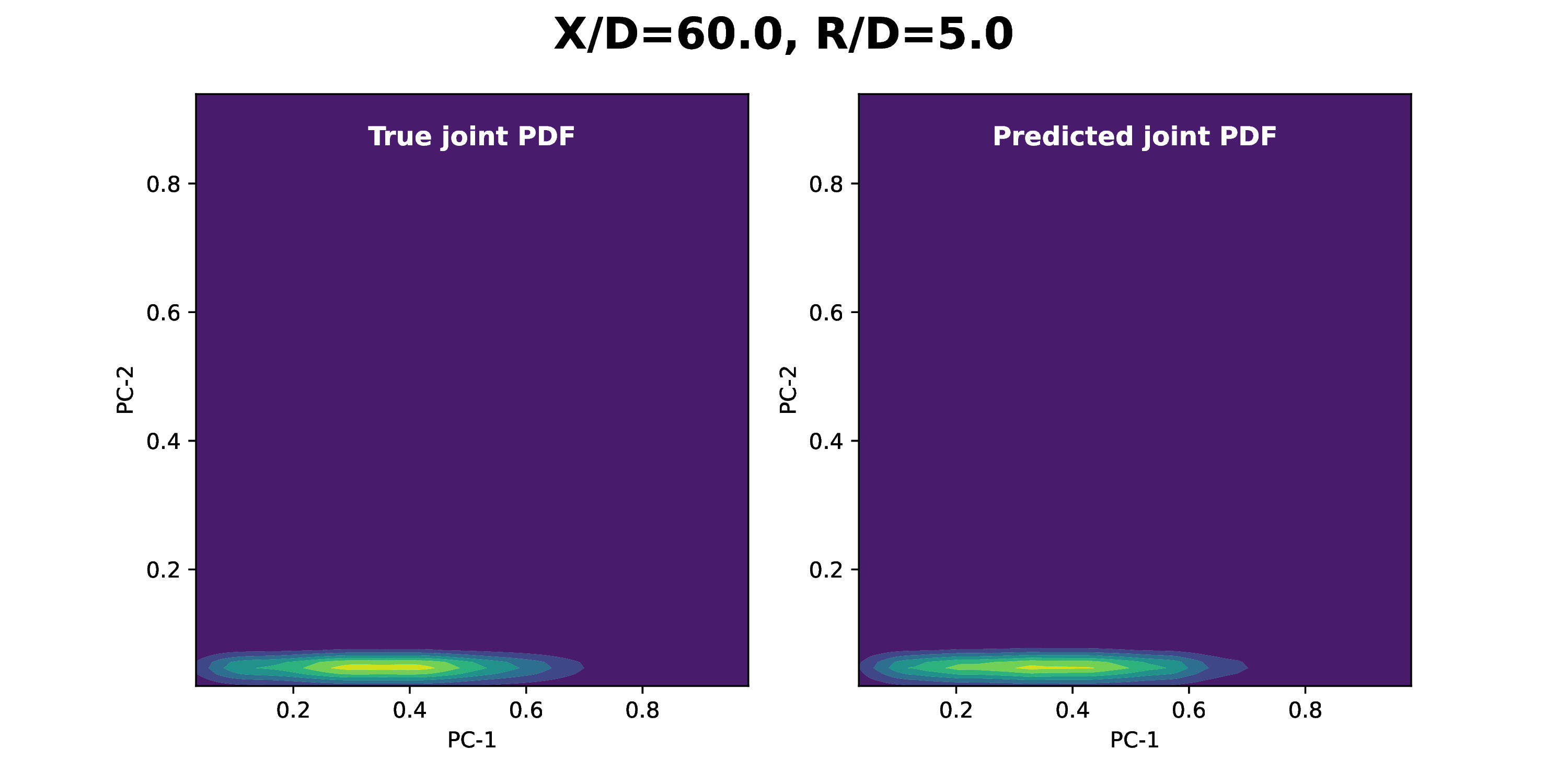}
\caption{\label{flamef} Comparison of contour plots between DeepONet vs KDE based joint PDFs for Sandia flame F}
\end{center}
\end{figure*}

The contour plots of joint PDFs are compared between DeepONet and KDE for the 3 Sandia flames, D, E and F in Figs. \ref{flamed}, \ref{flamee} and \ref{flamef}. It may be noted that the DeepONet was only trained on a few samples from flame E and used for prediction of joint PDFs on all $3$ flames. The joint PDFs are plotted at different axial and radial locations to show the evolution of the flame mixture through different non-equilibrium effects and that the different locations are characterized by different joint PDF shapes. These shapes are adequately captured by the DeepONet model that adopts as inputs their parameterization in terms of the unconditional means for the 2 PCs. Some errors may be observed in the joint PDFs reconstructed at certain locations of flame F. In future, we will try to address these errors by including more PCs in the construction of joint PDFs and by conditioning the DeepONet on the PC variances in addition to the means.

\section{Conclusions}

The objective of this work was to develop a generalizable model for learning joint PDFs in turbulent flames using the DeepONet network. The DeepONet based joint PDF approach was demonstrated on Sandia flames D, E and F, where the network was trained on data from flame E but predictions of joint PDFs were carried out on all 3 flames. The results show that the DeepONet based reconstruction approach can reasonably predict the joint PDF shapes for unseen turbulent flames and flame conditions. The results obtained are very encouraging and more work will be carried out on addressing inaccuracies and improving the generalizability of the DeepONet based joint PDF approach.

\section{Acknowledgements}
This work is supported by the National Science Foundation under grant no. 1941430.

\printbibliography

\end{document}